\documentstyle[12pt,fleqn]{article}

\font\twlgot =eufm10 scaled \magstep1
\font\egtgot =eufm8
\font\sevgot =eufm7

\font\twlmsb =msbm10 scaled \magstep1
\font\egtmsb =msbm8
\font\sevmsb =msbm7

\newfam\gotfam
\def\pgot{\fam\gotfam\twlgot}
\textfont\gotfam\twlgot
\scriptfont\gotfam\egtgot
\scriptscriptfont\gotfam\sevgot
\def\got{\protect\pgot}

\newfam\msbfam
\textfont\msbfam\twlmsb
\scriptfont\msbfam\egtmsb
\scriptscriptfont\msbfam\sevmsb
\def\Bbb{\protect\pBbb}
\def\pBbb{\relax\ifmmode\expandafter\Bb\else\typeout{You cann't use
Bbb in text mode}\fi}
\def\Bb #1{{\fam\msbfam\relax#1}}

\newcommand{\gO}{{\got O}}
\newcommand{\gQ}{{\got Q}}

\def\thebibliography#1{\section*{
References}\list
  {\arabic{enumi}.}{\settowidth\labelwidth{#1}\leftmargin\labelwidth
    \advance\leftmargin\labelsep
    \usecounter{enumi}}
    \def\newblock{\hskip .11em plus .33em minus .07em}
    \sloppy\clubpenalty4000\widowpenalty4000
    \sfcode`\.=1000\relax}

\let\Large=\large

\def\op#1{\mathop{\fam0 #1}\limits}

\newcommand{\di}{{\rm dim\,}}

\newcommand{\beq}{\begin{equation}}
\newcommand{\eeq}{\end{equation}}
\newcommand{\ben}{\begin{eqnarray}}
\newcommand{\een}{\end{eqnarray}}
\newcommand{\be}{\begin{eqnarray*}}
\newcommand{\ee}{\end{eqnarray*}}
\newcommand{\bea}{\begin{eqalph}}
\newcommand{\eea}{\end{eqalph}}

\newcommand{\cO}{{\cal O}}
\newcommand{\cQ}{{\cal Q}}

\newcommand{\la}{\lambda}
\newcommand{\La}{\Lambda}
\newcommand{\f}{\phi}

\newcommand{\m}{\mu}
\newcommand{\g}{\gamma}
\newcommand{\G}{\Gamma}

\newcommand{\th}{\theta}

\newcommand{\si}{\sigma}

\newcommand{\wh}{\widehat}
\newcommand{\ol}{\overline}
\newcommand{\dr}{\partial}

\newcommand{\ar}{\op\longrightarrow}

\newcounter{eqalph}
\newcounter{equationa}
\newcounter{example}
\newcounter{remark}
\newcounter{theorem}
\newcounter{proposition}
\newcounter{lemma}
\newcounter{corollary}
\newcounter{definition}

\def\theremark{\arabic{remark}}

\def\thedefinition{\arabic{definition}}

\newenvironment{prop}{\refstepcounter{definition} \medskip\noindent
PROPOSITION \thedefinition.\it}{\medskip }
\newenvironment{lem}{\refstepcounter{definition} \medskip\noindent  LEMMA
\thedefinition.\it }{\medskip }

\newenvironment{eqalph}{\stepcounter{equation}
\setcounter{equationa}{\value{equation}}
\setcounter{equation}{0}

\begin{eqnarray}}{\end{eqnarray}\setcounter{equation}{\value{equationa}}}

\begin{document}
\hbox{}

{\parindent=0pt 

{ \Large \bf Cohomology of the infinite-order jet space}
\bigskip

{\sc G.Giachetta$^\dagger$\footnote{giachetta@campus.unicam.it},
L.Mangiarotti$^\dagger$\footnote{mangiaro@camserv.unicam.it}  and
G.Sardanashvily$^\ddagger$}\footnote{sard@campus.unicam.it;
sard@grav.phys.msu.su} 

{ \small

{\it $^\dagger$ Department of Mathematics and Physics, University of Camerino,
62032 Camerino (MC), Italy

$^\ddagger$ Department of Theoretical Physics,
Physics Faculty, Moscow State University, 117234 Moscow, Russia}
\bigskip

{\bf Abstract} De Rham cohomology, $d_V$- and $d_H$-cohomology of the
differential algebra of locally pull-back exterior forms on the
infinite-order jet manifold of a smooth fibre bundle are calculated. 
\medskip

  } }

\section{Introduction}

Let $Y\to X$ be a smooth fibre bundle
(throughout the paper, smooth manifolds are assumed to be
real, finite-dimensional, Hausdorff,
paracompact, and connected).
We study cohomology of differential algebras of exterior forms on the
infinite-order jet space
$J^\infty Y$ of $Y\to X$. This cohomology plays an important role in some
physical  models, e.g., in the calculus of variations in Lagrangian field
theory [1-4] and in the field-antifield BRST formalism for constructing the
descent equations [5-8]. 

Recall that the infinite-order jet space of a smooth fibre bundle $Y\to X$ is
defined as a projective limit $(J^\infty Y,\pi^\infty_r)$ of the surjective
inverse system
\beq
X\op\longleftarrow^\pi Y\op\longleftarrow^{\pi^1_0}\cdots \longleftarrow
J^{r-1}Y \op\longleftarrow^{\pi^r_{r-1}} J^rY\longleftarrow\cdots 
\label{5.10}
\eeq
of finite-order jet manifolds $J^rY$ \cite{tak2,bau,tak1,book}. Provided with
the projective limit topology, $J^\infty Y$ is a paracompact Fr\'echet (but
not Banach)  manifold \cite{tak2,bau,abb}. Given a bundle coordinate chart
$(\pi^{-1}(U_X); x^\la,y^i)$ on the fibre bundle $Y\to X$, we have the
coordinate chart
$((\pi^\infty)^{-1}(U_X); x^\la, y^i_\La)$, $0\leq|\La|,$ on $J^\infty Y$,
together with the transition functions  
\beq
{y'}^i_{\la+\La}=\frac{\dr x^\m}{\dr x'^\la}d_\m y'^i_\La, \label{55.21}
\eeq
where $\La=(\la_k...\la_1)$, $|\La|=k$,
is a multi-index, 
 $\la+\La$ is the multi-index $(\la\la_k\ldots\la_1)$ and
$d_\la$ are the total derivatives 
\be
d_\la = \dr_\la + \op\sum_{|\La|=0} y^i_{\la+\La}\dr_i^\La.
\ee

In studying cohomology of differential algebras on the infinite-order jet
manifold, the key point lies in the following fact [3,4] (see Appendix A).

\begin{lem} A smooth fibre bundle
$Y$ is a strong deformation retract of its
infinite-order jet space $J^\infty Y$.
\end{lem}

 Since $J^\infty Y$ is
paracompact, it follows that there is an isomorphism 
\beq
H^*(J^\infty Y,\Bbb R)= H^*(Y,\Bbb R) \label{lmp80}
\eeq 
of the cohomology groups of
the infinite-order jet space $J^\infty Y$ with coefficients in
the constant sheaf $\Bbb R$ and those $H^*(Y,\Bbb R)$ of the fibre bundle $Y$. 

The goal is to construct a resolution of the constant sheaf $\Bbb R$ on
$J^\infty Y$ by acyclic sheaves of some differential algebra on $J^\infty Y$.
Then the well-known theorem on cohomology of global
sections of these sheaves  (\cite{hir}, Theorem 2.12.1) can be called into
play. For the sake of convenience, we will agree to call it the general De
Rham theorem.

Given the surjective inverse system (\ref{5.10}), we have the
direct system 
\beq
\gO^*_X\op\longrightarrow^{\pi^*} \gO^*_Y 
\op\longrightarrow^{\pi^1_0{}^*} \gO_1^*
\op\longrightarrow^{\pi^2_1{}^*} \cdots \op\longrightarrow^{\pi^r_{r-1}{}^*}
 \gO_r^* \longrightarrow\cdots \label{5.7'}
\eeq
of ringed spaces
$(J^rY,\gO^*_r)$  whose structure sheaves
$\gO^*_r$ are sheaves of differential $\Bbb R$-algebras of exterior
forms on finite-order jet manifolds $J^rY$, and
$\pi^r_{r-1}{}^*$ are the pull-back morphisms.
We follow the terminology of
Ref. \cite{hir} where by a sheaf is meant a sheaf bundle.  The
direct system (\ref{5.7'}) admits a direct limit
$\gQ^*_\infty$ which is 
a sheaf of differential exterior $\Bbb R$-algebras on
the infinite-order jet space
$J^\infty Y$. 

Accordingly, we have the direct system 
\beq
\cO^*(X)\op\longrightarrow^{\pi^*} \cO^*(Y) 
\op\longrightarrow^{\pi^1_0{}^*} \cO_1^*
\op\longrightarrow^{\pi^2_1{}^*} \cdots \op\longrightarrow^{\pi^r_{r-1}{}^*}
 \cO_r^* \longrightarrow\cdots \label{5.7}
\eeq
of the structure algebras $\cO^*_r=\G(J^rY,\gO^*_r)$ of global sections of
the sheaves $\gO^*_r$, i.e., $\cO^*_r$ are differential $\Bbb R$-algebras
of (global) exterior forms on finite-order jet manifolds
$J^rY$. The direct limit $(\cO^*_\infty,\pi^{\infty*}_r)$ of the direct system
(\ref{5.7}) is a differential $\Bbb R$-algebra of
all exterior forms on finite-order jet manifolds modulo the pull-back
identification. Therefore, one usually thinks of elements of $\cO^*_\infty$ as
being the pull-back onto $J^\infty Y$ of exterior forms on finite-order jet
manifolds. 

It should be emphasized that 
the direct limit $\cO^*_\infty$ of the direct system (\ref{5.7}) of
structure algebras of sheaves $\gO^*_r$ fails to coincide with the structure
algebra
$\cQ^*_\infty=\G(J^\infty Y,\gQ^*_\infty)$ of the direct limit
$\gQ^*_\infty$ of these sheaves. By definition,
$\gQ^*_\infty$ is the sheaf of germs of local exterior forms
on finite-order jet manifolds. These local forms constitute a
presheaf $\gO^*_\infty$  from which the sheaf $\gQ^*_\infty$ is
constructed. It means that, given a section
$\f\in\G(\gQ^*_\infty)$ of $\gQ^*_\infty$ over an open subset $U\in J^\infty Y$
and any point
$q\in U$, there exists a neighbourhood $U_q$ of $q$ such that
$\f|_{U_q}$ is the pull-back of a local exterior form on some finite-order jet
manifold. However, $\gO^*_\infty$ does not coincide with the canonical
presheaf $\G(U,\gQ^*_\infty)$ of sections of the sheaf $\gQ^*_\infty$.
There are obvious monomorphisms 
$\cO^*_\infty \to\cQ^*_\infty$ and  $\gO^*_\infty
\to\G(\gQ^*_\infty)$.

For short, we
agree to call $\gQ^*_\infty$ and $\cQ^*_\infty$ the sheaf and algebra 
of locally pull-back exterior forms on $J^\infty Y$. 
Being restricted to a coordinate chart $(\pi^\infty)^{-1}(U_X)$ on $J^\infty
Y$, elements of 
$\cQ^*_\infty$ 
can be written in the familiar coordinate form, where basic forms
$\{dx^\la\}$ and contact 1-forms
$\{\th^i_\La=dy^i_\La -y^i_{\la+\La}dx^\la\}$ provide 
the local generators of the algebra $\cQ_\infty^*$. There is
the canonical splitting of the space of $m$-forms
\be
\cQ^m_\infty =\cQ^{0,m}_\infty\oplus
\cQ^{1,m-1}_\infty\oplus\ldots\oplus\cQ^{m,0}_\infty
\ee
into spaces $\cQ^{k,m-k}_\infty$ of $k$-contact forms. 
Accordingly, the
exterior differential on $\cQ_\infty^*$ is
decomposed into the sum $d=d_H+d_V$
of horizontal and vertical differentials
\be
 d_H:\cQ_\infty^{k,s}\to \cQ_\infty^{k,s+1}, \qquad  d_V:\cQ_\infty^{k,s}\to
\cQ_\infty^{k+1,s}
\ee
which obey the nilpotency rule
\beq
d_H\circ d_H=0, \qquad d_V\circ d_V=0, \qquad d_V\circ d_H
+d_H\circ d_V=0. \label{lmp50}
\eeq

Traditionally, one has tried to 
introduce the algebra of locally pull-back forms on $J^\infty Y$ in a
standard geometric way
\cite{tak2,bau,tak1,abb}.  The difficulty lies in the geometric
interpretation of derivations of the $\Bbb R$-ring $\cQ^0_\infty$ of locally
pull-back functions as vector fields on the Fr\'echet manifold $J^\infty
Y$ and their duals as differential forms on
$J^\infty Y$. Therefore, one usually considers the 
subalgebra 
$\cO^*_\infty$ of pull-back exterior forms on $J^\infty Y$, given as the
direct limit of the direct system (\ref{5.7}). Accordingly,  the 
infinite-order De Rham complex of these forms 
 \beq
0\to \Bbb R\to
\cO^0_\infty\op\longrightarrow^d\cO^1_\infty\op\longrightarrow^d
\cdots
\label{5.13}
\eeq
is the direct limit of the De Rham complexes of exterior forms on
finite-order jet manifolds. Hence, as was repeatedly proved,
the cohomology groups $H^*(\cO^*_\infty)$ of the complex (\ref{5.13}) are equal
to the De Rham cohomology groups
$H^*(Y)$ of the fibre bundle $Y$ [1-4].  At the
same time, $d_H$- and $d_V$-cohomology of the differential algebra
$\cO^*_\infty$ remains unknown. The algebraic Poincar\'e lemma on the local
exactness of  differentials $d_H$ and $d_V$ only has been repeatedly proved
(see, e.g., [13-15]). 

Here, we show that the problem of cohomology of differential forms on the
infinite-order jet space $J^\infty Y$  has a comprehensive solution by enlarging
the algebra 
$\cO^*_\infty$ to the algebra $\cQ^*_\infty$. From the physical viewpoint, it
enables one also to study effective field theories whose Lagrangians involve
derivatives of arbitrary high order. The key point is that the Fr\'echet
manifold
$J^\infty Y$ admits a partition of unity performed by elements of the ring
$\cQ^0_\infty$ of locally pull-back functions on $J^\infty Y$ \cite{tak2,bau}.
It follows that the sheaf
$\gQ^*_\infty$ of $\cQ^0_\infty$-modules on
$J^\infty Y$ is fine and, consequently, acyclic. Therefore, studying different
resolutions performed by subsheaves of the sheaf $\gQ^*_\infty$, one may hope
to get a complete picture of cohomology of the differential algebra
$\cQ^*_\infty$. 

Given a smooth fibre bundle $Y\to X$, we will show the following. 
\begin{itemize}\begin{enumerate}
\item The De Rham cohomology groups of the differential algebra
$\cQ^*_\infty$ are isomorphic to those of the fibre bundle $Y$.
\item Its $d_V$-cohomology groups are related to the cohomology groups
of the fibre bundle $Y$ with coefficients in the sheaf $\gO^*_X$ of
exterior forms on the base $X$. 
\item The $d_H$-cohomology groups of contact elements $\f\in
\cQ^{0<,*}_\infty$ of the algebra $\cQ^*_\infty$ are trivial.
\item In degrees $r<n= \di X$, the $d_H$-cohomology groups of its horizontal
elements $\f\in\cQ^{0,*}_\infty$ coincide with the De Rham cohomology groups of
the fibre bundle
$Y$.
\end{enumerate}\end{itemize}

Note that, as was mentioned above, the result (i) is also true for the
differential algebra
$\cO^*_\infty$. The result (iii) recovers that in Refs. \cite{ander0,ander},
obtained by means of the Mayer-Vietoris sequence.

Point out the  particular case of an affine bundle $Y\to X$, interesting
for physical applications, e.g., to BRST theory. In this case, $X$ is a
strong deformation retract of $Y$ and the cohomology of $Y$ under
consideration is equal to that of $X$. Then the above results are
reformulated as follows.
\begin{itemize}\begin{enumerate}
\item Any closed form $\f\in\cQ^*_\infty$ is decomposed into the sum
$\f=\varphi + d\xi$ where $\varphi\in\cO^*(X)$ is a closed form on the base
$X$.
\item Any $d_V$-closed form $\f\in\cQ^*_\infty$ is $d_V$-exact.
\item Any $d_H$-closed form $\f\in\cQ^*_\infty$ is decomposed into the sum
$\f=\varphi + d_H\xi$ where $\varphi\in\cO^*(X)$ is a closed form on the base
$X$.
\end{enumerate}\end{itemize}

The results (i) and (ii) are also true for the
differential algebra $\cO^*_\infty$.

\section{De Rham cohomology}

Let us start from De Rham cohomology of the differential algebra
$\cQ^*_\infty$ of locally pull-back exterior forms on the infinite-order jet
manifold $J^\infty Y$.  We consider the complex of sheaves of
$\cQ^0_\infty$-modules
 \beq
0\to \Bbb R\to
\gQ^0_\infty\op\longrightarrow^d\gQ^1_\infty\op\longrightarrow^d
\cdots
\label{lmp71}
 \eeq
on the infinite-order jet space $J^\infty Y$ and the corresponding 
infinite-order De Rham complex
 \beq
0\to \Bbb R\to
\cQ^0_\infty\op\longrightarrow^d\cQ^1_\infty\op\longrightarrow^d
\cdots
\label{5.13'}
\eeq
of algebras of locally pull-back exterior forms on $J^\infty Y$. 

Since locally pull-back
exterior forms fulfill the Poincar\'e lemma, the complex of sheaves
(\ref{lmp71}) is exact. Since the paracompact space $J^\infty Y$ admits a
partition of unity performed by elements of $\cQ^0_\infty$, the
sheaves
$\gQ^r_\infty$ of
$\cQ^0_\infty$-modules are fine for all $r\geq 0$. Then
they are acyclic, i.e., the cohomology groups $H^{>0}(J^\infty
Y,\gQ^r_\infty)$ of the paracompact space $J^\infty Y$ with coefficients in
sheaves
$\gQ^r_\infty$ vanish. Consequently, the exact sequence
(\ref{lmp71}) is a fine resolution of the constant sheaf $\Bbb R$ on
$J^\infty Y$. Then, in accordance with the above mentioned general De
Rham theorem, we have an isomorphism
\beq
H^*(\cQ^*_\infty)=H^*(J^\infty Y,\Bbb R) \label{lmp82}
\eeq 
of the De Rham cohomology groups $H^*(\cQ^*_\infty)$ of the
differential algebra $\cQ^*_\infty$ and the cohomology groups
$H^*(J^\infty Y,\Bbb R)$ of the infinite-order jet space $J^\infty Y$ with
coefficients in the constant sheaf $\Bbb R$.
Combining isomorphisms (\ref{lmp80}) and (\ref{lmp82}), we come to the
manifested assertion.

\begin{prop} There is an isomorphism
\be
H^*(\cQ^*_\infty)=H^*(Y) 
\ee
of the De Rham cohomology groups $H^*(\cQ^*_\infty)$ 
of the differential algebra
$\cQ^*_\infty$  to the De Rham cohomology
groups
$H^*(Y)$ of the fibre bundle $Y$.
\end{prop}

\section{Cohomology of $d_V$}

Due to the nilpotency rule (\ref{lmp50}), the vertical and horizontal
differentials $d_V$ and $d_H$ define the bicomplex of sheaves
\beq
\begin{array}{ccccrlcrlccrlccrlcc}
& & & & _{d_V} & \put(0,-10){\vector(0,1){20}} & & _{d_V} &
\put(0,-10){\vector(0,1){20}} & &  & _{d_V} &
\put(0,-10){\vector(0,1){20}} & & &  _{d_V} &
\put(0,-10){\vector(0,1){20}}& &\\ 
 &  & 0 & \to & &\gQ^{k,0}_\infty &\ar^{d_H} & & \gQ^{k,1}_\infty &
\ar^{d_H} &\cdots  & & \gQ^{k,m}_\infty &\ar^{d_H} &\cdots & &
\gQ^{k,n}_\infty & & \\  
 & &  &  & & \vdots & & & \vdots  & & & 
&\vdots  & & & &
\vdots & &  \\ 
& & & & _{d_V} &\put(0,-10){\vector(0,1){20}} & & _{d_V} &
\put(0,-10){\vector(0,1){20}} & & &  _{d_V}
 & \put(0,-10){\vector(0,1){20}} & &  & _{d_V} & \put(0,-10){\vector(0,1){20}}
 & &\\
0 & \to & \Bbb R & \to & & \gQ^0_\infty &\ar^{d_H} & & \gQ^{0,1}_\infty &
\ar^{d_H} &\cdots  & &
\gQ^{0,m}_\infty & \ar^{d_H} & \cdots & &
\gQ^{0,n}_\infty & &\\
& & & & _{\pi^{\infty*}}& \put(0,-10){\vector(0,1){20}} & & _{\pi^{\infty*}} &
\put(0,-10){\vector(0,1){20}} & & &  _{\pi^{\infty*}}
 & \put(0,-10){\vector(0,1){20}} & &  & _{\pi^{\infty*}} &
\put(0,-10){\vector(0,1){20}} & & \\
0 & \to & \Bbb R & \to & & \gO^0_X &\ar^d & & \gO^1_X &
\ar^d &\cdots  & &
\gO^m_X & \ar^d & \cdots & &
\gO^n_X & \ar^d & 0\\
& & & & &\put(0,-10){\vector(0,1){20}} & & &
\put(0,-10){\vector(0,1){20}} & & & 
 & \put(0,-10){\vector(0,1){20}} & & &   &
\put(0,-10){\vector(0,1){20}} & & \\
& & & & &0 & &  & 0 & & & & 0 & & & & 0 & & 
\end{array}
\label{7}
\eeq
[1-4,10,13-15]. The rows and columns of these bicomplex are horizontal and
vertical complexes. 
Moreover, the above mentioned algebraic
Poincar\'e lemma is obviously extended to elements of $\cQ^*_\infty$ and 
leads to the following.

\begin{lem}
The columns and rows of the bicomplex (\ref{7}) are exact sequences of
sheaves.
\end{lem}

It follows that, since all sheaves $\gQ^{k,m}_\infty$ are fine, the columns
and rows of the bicomplex (\ref{7}) are fine resolutions of their first terms.
Then the general De Rham theorem on a resolution of a sheaf on a
paracompact manifold can be used again.

Let us consider a vertical exact sequence of sheaves 
\beq
0\to \gO^m_X \ar^{\pi^{\infty*}} \gQ^{0,m}_\infty \ar^{d_V}\cdots \ar^{d_V} 
\gQ^{k,m}_\infty \ar^{d_V}\cdots, \qquad m\leq n, \label{lmp90'}
\eeq
and the corresponding complex of their structure algebras
\beq
0\to \cQ^m(X) \ar^{\pi^{\infty*}} \cQ^{0,m}_\infty \ar^{d_V}\cdots \ar^{d_V} 
\cQ^{k,m}_\infty \ar^{d_V}\cdots, \qquad m\leq n. \label{lmp90}
\eeq
The exact sequence (\ref{lmp90'}) is a resolution of the sheaf
$\pi^{\infty*}\gO^m_X$ of the pull-back onto $J^\infty Y$ of exterior forms on
the base $X$. Then, by virtue of the general De Rham theorem, we have an
isomorphism
\beq
H^*(d_V,m)=H^*(J^\infty Y,\pi^{\infty*}\gO^m_X) 
\eeq
of the cohomology groups $H^*(d_V,m)$ of the complex of
differential algebras (\ref{lmp90}) to the cohomology groups $H^*(J^\infty
Y,\pi^{\infty*}\gO^m_X)$ of the infinite-order jet manifold $J^\infty Y$ with
coefficients in the sheaf
$\pi^{\infty*}\gO^m_X$. Since $J^\infty Y$ and $Y$ are homotopic, there is an
isomorphism of cohomology groups
\be
H^*(J^\infty Y,\pi^{\infty*}\gO^m_X)=H^*(Y,\pi^*\gO^m_X).
\ee
Thus, it is stated the following.

\begin{prop} There is an isomorphism of the cohomology groups
\be
H^*(d_V,m)=H^*(Y,\pi^*\gO^m_X).
\ee
\end{prop}

In particular, if $Y\to X$ is an affine bundle, we have
\be
H^*(d_V,m)=H^*(Y,\pi^*\gO^m_X)= H^*(X,\gO^m_X)=0
\ee
because the sheaf $\gO^m_X$ on $X$ is fine. This result also follows directly
from the expression for the corresponding homotopy operator and, therefore,
remains true for to the differential algebra $\cO^*_\infty$ (see Appendix B).

\section{Cohomology of $d_H$}

Turn now to the rows of the bicomplex (\ref{7}) (excluding the bottom one
which is obviously the De Rham complex on the base $X$). We have the exact
sequences of sheaves 
\ben
&& 0\to \gQ^{k,0}_\infty \ar^{d_H}\gQ^{k,1}_\infty\ar^{d_H}\cdots  
\op\longrightarrow^{d_H} 
\gQ^{k,n}_\infty, \qquad k>0, \label{lmp91}\\
&& 0\to \Bbb R \to \gQ^0_\infty \ar^{d_H} \gQ^{0,1}_\infty
\ar^{d_H}\cdots  
\op\longrightarrow^{d_H} 
\gQ^{0,n}_\infty .\label{lmp92'}
\een
Speaking rigorously, the exact sequences
(\ref{lmp91}) and (\ref{lmp92'}) fail to be fine resolutions of the sheaves
$\gQ^{k,0}_\infty$ and $\Bbb R$, respectively, because of their last terms. At
the same time, following  directly the proof of the above mentioned Theorem
2.12.1 in Ref. \cite{hir} till these terms, one
can show the following.

\begin{prop} \label{lmp99'} The cohomology
groups $H^r(d_H, k)$, $r<n$, of the complex   
\be
 0\to \cQ^{k,0}_\infty \ar^{d_H}\cQ^{k,1}_\infty\ar^{d_H}\cdots  
\op\longrightarrow^{d_H} 
\cQ^{k,n}_\infty
\ee
are
isomorphic to the cohomology groups $H^r(J^\infty Y,\gQ^{k,0}_\infty)$ of
$J^\infty Y$ with coefficients in the sheaf $\gQ^{k,0}_\infty$ and,
consequently, are trivial because the sheaf $\gQ^{k,0}_\infty$ is fine.
\end{prop}

\begin{prop} \label{lmp99} 
The cohomology groups $H^r(d_H)$, $r<n$, of the complex
\be
0\to \Bbb R \to \cQ^0_\infty \ar^{d_H} \cQ^{0,1}_\infty
\ar^{d_H}\cdots  
\op\longrightarrow^{d_H} 
\cQ^{0,n}_\infty
\ee
are isomorphic to the
cohomology groups $H^r(J^\infty Y,\Bbb R)$ of $J^\infty Y$ with
coefficients in the constant sheaf $\Bbb R$ and, consequently, to the
cohomology groups $H^r(Y,\Bbb R)$.
\end{prop}

Note that one can also study the exact sequences of presheaves
\be
&& 0\to \gO^{k,0}_\infty \ar^{d_H}\gO^{k,1}_\infty\ar^{d_H}\cdots  
\op\longrightarrow^{d_H} 
\gO^{k,n}_\infty, \qquad k>0,\\
&& 0\to \Bbb R \to \gO^0_\infty \ar^{d_H} \gO^{0,1}_\infty
\ar^{d_H}\cdots  
\op\longrightarrow^{d_H} 
\gO^{0,n}_\infty ,
\ee
but comes again to the results of Propositions \ref{lmp99'}, \ref{lmp99}.
Because $J^\infty Y$ is paracompact, the cohomology groups $H^*(J^\infty Y,
\gQ^{*,*}_\infty)$ of
$J^\infty Y$ with coefficients in a sheaf $\gQ^{*,*}_\infty$ and those
$H^*(J^\infty Y, \gO^{*,*}_\infty)$ with coefficients in a presheaf
$\gO^{*,*}_\infty$ are isomorphic. It follows that the cohomology group
$H^0(J^\infty Y, \gO^{*,*}_\infty)$ of a
presheaf $\gO^{*,*}_\infty$ is isomorphic to the $\Bbb R$-module
$\cQ^{*,*}_\infty=H^0(J^\infty Y, \gQ^{*,*}_\infty)$, but not
$\cO^{*,*}_\infty$.

\section{Appendix A}

Here, we construct a desired  homotopy from the infinite order jet manifold 
$J^\infty Y$ to a fibre bundle 
$Y$ in an explicit form. Given a coordinate chart 
$((\pi^\infty)^{-1}(U_X); x^\la, y^i_\La)$ on $J^\infty Y$,
let us
consider the map
\be
&&[0,1]\times J^\infty Y\ni (t; x^\la, y^i, y^i_\La) \to (x^\la,
y^i,y'^i_\La)\in J^\infty Y, \qquad 0<|\La|,\\
&& y'^i_\La= f_k(t)y^i_\La +(1-f_k(t))\G_{(k)}{}^i_\La), \qquad |\La|=k>0,
\ee
where $\G_{(k)}$ is a section of the affine jet bundle $J^kY\to J^{k-1}Y$ and 
$f_k(t)$ is a continuous (smooth) real function on $[0,1]$ such that
\be
f_k(t)=\left\{
\begin{array}{ll}
0, & \quad t\leq 1-2^{-k},\\
1, & \quad t\geq 1-2^{-(k+1)}.
\end{array}\right. 
\ee
A glance at the transition functions (\ref{55.21}) shows that, given in a
coordinate form, this map is well-defined. 
It is a desired homotopy from $J^\infty Y$ to $Y$ which is identified
with its image under the global section 
\be
\g: Y\to J^\infty Y, \qquad y^i_\La\circ \g =\G_{(k)}{}^i_\La
\circ\G_{(k-1)}\circ \cdots \circ \G_1, \quad |\La|=k.
\ee

\section{Appendix B}

We start from a remark that,
Studying cohomology of exterior forms on the infinite-order jet space
of an affine bundle, we can restrict our consideration  to
vector bundles
$Y\to X$  without loss of generality as follows.   Let $Y\to X$ be a smooth
affine bundle modelled over a smooth vector bundle
$\ol Y\to X$. 
A glance at the transformation law (\ref{55.21}) shows that $J^\infty Y\to X$
is an affine topological bundle modelled on the vector bundle $J^\infty
\ol Y\to X$. This affine bundle admits a global section
$J^\infty s$ which is the infinite-order jet prolongation of a global section
$s$ of
$Y\to X$. With $J^\infty s$, we have a homeomorphism
\be
\wh s_\infty: J^\infty Y \ni q \mapsto q -(J^\infty s)(\pi^\infty(q))\in
J^\infty\ol Y
\ee
of the topological spaces $J^\infty Y$ and $J^\infty\ol Y$,
together with an exterior algebra isomorphism
$\wh s_\infty^*:
\ol\cO^*_\infty\to\cO^*_\infty$.
Moreover, it is readily observed that the pull-back morphism $\wh
s_\infty^*$ commutes with the differentials $d$, $d_V$ and $d_H$.
Therefore, the differential algebras 
$\cQ^*_\infty$ and $\ol\cQ^*_\infty$ (as like as $\cO^*_\infty$ and
$\ol\cO^*_\infty$) have the same $d$-, $d_V$- and $d_H$-cohomology. 

Let $Y\to X$ be a vector bundle. Let us consider the vertical complex 
\be
0\to \cO^m(X) \ar^{\pi^{\infty*}} \cO^{0,m}_\infty \ar^{d_V}\cdots \ar^{d_V} 
\cO^{k,m}_\infty \ar^{d_V}\cdots, \qquad m\leq n,
\ee
of differential algebras of pull-back exteriror forms on $J^\infty Y$.
Its local exactness on a coordinate chart
$((\pi^\infty)^{-1}(U_X); x^\la, y^i_\La)$, $0\leq|\La|,$ on $J^\infty Y$
follows from a version of the Poincar\'e lemma with
parameters (see, e.g., \cite{tul}). We have the corresponding 
homotopy operator
\be
\si=\int^1_0 t^k [ \ol y\rfloor\f(x^\la, ty^i_\la)]dt, \qquad \f\in
\c0^{k,m}_\infty,
\ee
where $\ol y=y^i_\La\dr_i^\La$. Since $Y\to X$ is a
vector bundle, it is readily observed that, given in a coordinate form, this
homotopy operator is globally defined on $J^\infty Y$, and so is the exterior
form $\si$.

\end{document}